\documentclass[11pt,superscriptaddress,aps,prd,preprint]{revtex4}
\usepackage{amsmath}

\makeatletter
\usepackage[T1]{fontenc}
\usepackage{amsmath}
\usepackage{amssymb}
\usepackage{graphicx}

\usepackage{color}

\usepackage{slashed}

\DeclareSymbolFont{extraitalic}      {U}{zavm}{m}{it}
\DeclareMathSymbol{\Qoppa}{\mathord}{extraitalic}{161}
\DeclareMathSymbol{\qoppa}{\mathord}{extraitalic}{162}
\DeclareMathSymbol{\Stigma}{\mathord}{extraitalic}{167}
\DeclareMathSymbol{\Sampi}{\mathord}{extraitalic}{165}
\DeclareMathSymbol{\sampi}{\mathord}{extraitalic}{166}
\DeclareMathSymbol{\stigma}{\mathord}{extraitalic}{168}

\newcommand{\bea}{\begin{eqnarray}}
\newcommand{\eea}{\end{eqnarray}}

\begin{document}

\title{ A study on causality in $f(R,\phi,X)$ theory}

\author{J. S. Gon\c{c}alves}\email[]{junior@fisica.ufmt.br }
\affiliation{Instituto de F\'{\i}sica, Universidade Federal de Mato Grosso,\\
78060-900, Cuiab\'{a}, Mato Grosso, Brazil}

\author{A. F. Santos}\email[]{alesandroferreira@fisica.ufmt.br}
\affiliation{Instituto de F\'{\i}sica, Universidade Federal de Mato Grosso,\\
78060-900, Cuiab\'{a}, Mato Grosso, Brazil}

\begin{abstract}

The $k$-essence modified $f(R)$ gravity model, i.e., $f(R,\phi,X)$ theory is studied. The question of violation of causality, in the framework of G\"{o}del-type universes, is investigated in this gravitational model. Causal and non-causal solutions are allowed. A critical radius for non-causal solution is calculated. It is shown that the violation of causality depends on the content of matter.

\end{abstract}

\maketitle

\section{Introduction}

The discovery of accelerated cosmic expansion \cite{expansao_ace,expansao_ace2, Riess} led to the search of theoretical models  consistent with observational data \cite{fr1}. In this search, two possibilities have been developed to explain this phenomenon: (i) alternative models that modify the General Relativity (GR) and (ii) add an exotic component, called dark energy, in GR. For a review, see \cite{darkenergy}. The reasons and motivations that lead to alternatives gravity theories have changed over the years. Some models are based on theoretical reasons while others are more phenomenological, among other considerations. There are a number of theoretical models that were developed to study the accelerated expansion of the universe. Here our main interesting is in the theories of gravity: $f(R)$, $k$-essence and $k$-essence modified $f(R)$ gravity.

The $f(R)$ gravity is a modified gravitational theory in which the standard Einstein-Hilbert action is replaced by an arbitrary function of the Ricci scalar \cite{fr2, fr3}.  These theories have received considerable attention in the last years motivated by the fact that they can explain the observed accelerating late expansion of the universe \cite{Alcaniz, Capo}. There are several studies on $f(R)$ gravity, such as, solar system tests \cite{Chiba, Amen}, Newtonian limit \cite{Dicke, Soti, limite_newton}, gravitational stability \cite{Fara, estabilidade}, singularities \cite{singularidade}, energy conditions have been used to place constraints on the theory \cite{Kung, Santos}, cosmological perturbations during the inflationary epoch \cite{Chera}, gravitoelectromagnetism formalism in the context of metric $f(R)$ theory \cite{Dass}, among others.

Another interesting modified gravity model is the $k$-essence gravity. It emerged as an alternative dynamical dark energy model that has been developed in the context of inflationary theory \cite{k_inflation, k_inflation2}. The $k$-essence gravity also has been used with the objective to explain the accelerated expansion of the universe \cite{k_essece_1, kessece2, kessece3, kessece4}. The main idea of this theory is to minimally couple a scalar field with the gravity that depends only on the kinetic terms and does not depend on the scalar field itself. There are numerous applications with this theory. For example, the FRW $k$-essence cosmologies has been analyzed \cite{Agui}, exact solutions in $k$-essence for isotropic cosmology have been studied \cite{Pimentel},  dark matter to dark energy transition in $k$-essence cosmologies has been examined \cite{Forte}, power-law expansion in $k$-essence cosmology has been investigated \cite{Feins}, quantum cosmology with $k$-essence has been discussed \cite{Fabris}, G\"{o}del-type universe in $k$-essence gravity has been studied \cite{our}, among others. 

In this paper, the approach is to study a model that is a combination of both modifications, the $f(R)$ gravity and the $k$-essence theory,  called $f(R,\phi,X)$ theory, where $\phi$ is a scalar field and $X = \frac{1}{2}\partial^\mu\phi\partial_\nu\phi$. This model has been investigated in recent works such as \cite{kessece4, model1, model2}. So our main aim in this paper is to investigate the question of causality in this $k$-essence modified $f(R)$ gravitational model. To achieve this goal, the G\"{o}del universe and its generalizations will be considered.

In 1949,  Kurt G\"{o}del proposed an exact solution of Einstein equations for a homogeneous rotating universe. This solution leads to the possibility of Closed Timelike Curves (CTCs), which allow violation of causality and makes time-travel theoretically possible in this space-time \cite{godel}. CTCs are also admitted in other cosmological models, such as Kerr black hole, Van-Stockum model, cosmic string, among others \cite{ctc1,ctc2}. A generalization of the G\"{o}del solution, known as G\"{o}del-type solution, has been developed \cite{tipo_godel1}.  The G\"{o}del-type solutions add more details to the problem of causality. In addition, in the G\"{o}del-type universes there is a possibility of a causal rather than non-causal solution. These regions are determined from free parameters of the metric and limited by a critical radius $r_c$. From this critical radius there is violation of causality.

The problem on violation of causality has been investigated in several modified gravity theories, such as,  in  $f(R)$ theory \cite{fr_and_godel},  in $k$-essence theory \cite{our}, in Chern-Simons gravity \cite{chersimon_and_godel1, chersimon_and_godel2}, in $f(T)$ gravity \cite{ft_and_godel}, in $f(R,T)$ gravity \cite{frt_and_godel}, in bumblebee gravity \cite{bumblebeee_and_godel}, in Horava-Lifshitz gravity \cite{horava_and_godel}, Brans-Dicke theory \cite{brans_and_godel}, in $f(R,Q)$ gravity \cite{frq_and_godel}, among others. Here the consistency of G\"{o}del-type solutions and their implications about causality are analyzed in the $k$-essence modified $f(R)$ gravity.

This paper is organized as follows. In section 2, a brief introduction to $f(R)$ gravity and $k$-essence theory is presented. The $f(R,\phi,X)$ theory is introduced.  In sections 3 and 4, the G\"{o}del and G\"{o}del-type universes are analyzed in the $f(R,\phi,X)$ theory for different matter contents such as perfect fluid, perfect fluid plus electromagnetic field and only for electromagnetic field. In all cases, the question of causality is investigated. In section 5, some concluding remarks about the causal and non-causal solutions are discussed.

\section{Modified Gravity Models}

In this section, the $k$-essence modified $f(R)$ gravity is discussed. First, the $f(R)$ and $k$-essence gravity models are briefly introduced.

\subsection{$f(R)$ theory} 

$f(R)$ is a gravitational theory that generalizes GR. This generalization consist in replace the Ricci scalar $R$ in the Einstein-Hilbert action by a general function of $R$. The action that describes the $f(R)$ theory is 
\bea
     S = \int d^{4} x \sqrt{-g} \ \left(\frac{f(R)}{2 \beta} + \mathcal{L}_{m} \right),
\eea
where $\beta \equiv 8 \pi G$, $g$ is the determinant of the metric $g_{\mu\nu}$ and $\mathcal{L}_{m}$ is the lagrangian that describes the content of matter.  Varying the action with respect to the metric, after some manipulations, the field equations are obtained as
\begin{equation}
    R_{\mu\nu} f'_R - \left(\nabla_\mu\nabla_\nu -g_{\mu\nu} \Box\right) f'_R - \frac{1}{2} g_{\mu\nu} f =\beta\, T_{\mu\nu},\label{fe1}
\end{equation}
where $f = f(R)$, $f'_R \equiv df(R)/d(R)$, $\nabla_\mu$ is the covariant derivative,   $ \Box \equiv g^{\alpha\beta}\nabla_\alpha\nabla_\beta$ and $T_{\mu\nu}$ represents the energy-momentum tensor associated to the matter defined as 
\bea
T_{\mu\nu} = -\frac{2}{\sqrt{-g}}\frac{\delta \left(\sqrt{-g}\mathcal{L}_{m} \right)}{ \delta g^{\mu\nu}}.\label{EM}
\eea

A constraint, often used to simplify the field equations, comes from the trace of eq. (\ref{fe1}), which is given by
\bea
 f'_R R -2f + 3\Box f'_R = \beta T, 
\eea
where $T=g^{\mu\nu}T_{\mu\nu}$  is the trace of the energy-momentum tensor. Then the relation between $R$ and $T$ differs from the trace of the Einstein equation, i.e., $R=-\beta T$. This is an indication that the field equations of $f(R)$ theories will admit a larger variety of solutions than GR.

\subsection {$k$-essence theory}

The $k$-essence theory is characterized by a scalar field with a non-canonical kinetic energy. In this model a single scalar field $\phi$ interacts with gravity through non-standard
kinetic terms. The action that describes the $k$-essence theory is given by
\begin{equation}\label{mod2}
     S = \int d^{4} x \sqrt{-g}  \left(\frac{R}{2\beta} - K(\phi,X)  + \mathcal{L}_{m} \right),
\end{equation}
where $K(\phi,X)$ is a function of the scalar field $\phi$ and its derivatives with $X=\partial^\mu\phi\partial_\mu\phi$. Here the scalar field depends only on time, i.e. $\phi=\phi(t)$.  

Varying the action with respect to the metric $g_{\mu\nu}$, the field equations are given as
\begin{equation}\label{ap1}
    \frac{1}{\beta} \left(R_{\mu\nu} - \frac{1}{2} g_{\mu\nu} R \right) = K g_{\mu\nu} - K_x \ \partial_\mu \phi \partial_\nu \phi + T_{\mu\nu},
\end{equation}
where, $K=K(\phi,X)$, $K_x = \frac{\partial K(\phi,X)}{\partial X}$ and $T_{\mu\nu}$ is the energy momentum tensor associated with the content of matter, as defined in eq. (\ref{EM}). In limit $K(\phi,X) \rightarrow 0$, GR is recovered. 

The second field equation is obtained varying the action with respect to the scalar field. Then 
\begin{equation}
    - K_{\phi}(\phi,X) + 2\nabla_{\mu} \left(K_{x}(\phi,X) \partial^{\mu} \phi \right)=0,\label{phi}
\end{equation}
with $K_\phi(\phi,X) = \frac{\partial K(\phi,X)}{\partial \phi}$ and $\nabla_{\mu}$ being the covariant derivative.

\subsection{$f(R,\phi,X)$ theory} 

In this section, a hybrid model, composed of $k$-essence and $f(R)$ gravity theories, is proposed. Here the main idea is to study the question of causality in this gravitational model, known as $f(R,\phi,X)$ theory. A general class of $k$-essence $f(R)$ gravity models is described by the action
\begin{equation}\label{mod12}
     S = \int d^{4} x \sqrt{-g} \ \left(\frac{f(R)}{2\beta} -2\Lambda - K(\phi,X) +  \mathcal{L}_{m}, \right),
\end{equation}
where $\Lambda$ is the cosmological constant. It is important to note that, the question about ghost degrees of freedom in this hybrid theory has been investigated and the no-ghost constraints on a special class of $k$-essence $f(R)$ gravity models have been found \cite{model1}.

By varying the action with respect to the metric, the field equations for the hybrid model are
\begin{equation}\label{mod3}
    \frac{1}{\beta} \left[f'_R R_{\mu\nu} - \left(\nabla_\mu\nabla_\nu -g_{\mu\nu} \Box\right) f'_R - \frac{1}{2} g_{\mu\nu} f - g_{\mu\nu}\Lambda \right] = -Kg_{\mu\nu} + 2K_x\partial_\mu \phi \partial_\nu\phi + T_{\mu\nu}. 
\end{equation}

In order to simplify the field equations, a constraint is imposed by the trace of eq. (\ref{mod3}) that is given as
\begin{equation}
    f'_R R -2f + 3\Box f'_R = \beta\left[-4K + 2K_x \partial^\alpha \phi \partial_\alpha \phi + T \right].
\end{equation}
Note that, it is different from the results of GR and $f(R)$. Therefore, the $f(R,\phi,X)$ gravity theory will admit a larger variety of solutions than GR, $f(R)$ and $k$-essence theory separately. In the next sections, the compatibility of $f(R,\phi,X)$ gravity theory with G\"{o}del and G\"{o}del-type universes as well as the question of causality are investigated.

\section{G\"{o}del universe in $f(R,\phi,X)$ theory}

The line element that represent the G\"{o}del universe is given by
\begin{equation}\label{godel_metric}
    ds^2 = a^2 \left(dt^2 - dx^2 + \frac{e^{2x}}{2}dy^2 - dz^2 + 2e^xdt \ dy \right),
\end{equation}
where $a$ is a positive number.  The non-zero Ricci tensor components are
\begin{align}
    R_{00} = 1, \ \ \ R_{02} = R_{20} = e^x, \ \ \ R_{22} = e^{2x},
\end{align}
and the Ricci scalar is
\begin{equation}
    R = \frac{1}{a^2}.
\end{equation}

By considering that the energy-momentum tensor is given as
\begin{equation}
    T_{\mu\nu} = \rho u_\mu u_\nu,
\end{equation}
with $\rho$ being the energy density of the fluid of matter, and $u$ being a unit time-like vector whose explicit contravariant components look like $u_\mu= (a,0,ae^x,0)$, the field equations, eq. (\ref{mod3}), become
\begin{align}
    \beta \left(2K_x\dot{\phi}^2 - a^2K + a^2\rho \right) - f_R' +\frac{a^2}{2} f + a^2\Lambda &= 0,\\
    \beta K - \frac{1}{2}f - \Lambda &= 0,\\
    \beta \left(\rho - K \right) - \frac{1}{a^2} f_R' + \frac{1}{2} f + \Lambda &=0,\\
    \beta \left(\rho -\frac{1}{2}K \right) - \frac{1}{a^2}f_R' + \frac{1}{4} f + \frac{1}{2} \Lambda &=0.
\end{align}
Here it has been assumed that the field $\phi$ is a function in the time coordinate only. This set of equations is satisfied if 
\begin{equation}
    \rho = \frac{1}{a^2\beta}f_R', \ \ \ \ \ \Lambda = \beta K -\frac{1}{2} f.
\end{equation}
In addition, a restriction on the $K(\phi, X)$ function is imposed, i.e., $\frac{\partial K(\phi, X)}{\partial X}=0$. Then the $K(\phi, X)$ function should only depend on $\phi$. However, the scalar field equation of motion, eq. (\ref{phi}), it implies $dK/d \phi =0$. Then $K$ must be a constant independent of $\phi$, which just shifts the cosmological constant by a given constant.
This result implies that the G\"{o}del metric solves the modified Einstein equations if and only if these conditions are satisfied. Therefore, in this case, the $k$-essence function turns out to be trivial, i.e., the $k$-essence $f(R)$ theory is reduced to the usual $f(R)$ gravity. For the case $f = R$, $f' = 1$ the G\"{o}del universe in GR is recovered. This implies that Closed Time-like Curves (CTC) are possible and, as a consequence, violation of causality is permitted. 

In the next section, a generalization of the G\"{o}del solution will be considered and the question of causality will be investigated for different contents of matter.

\section{G\"{o}del-type universe in $f(R,\phi,X)$ theory}

In order to obtain more information on the issue of causality, a generalization of the G\"{o}del solution, called G\"{o}del-type metrics, has been proposed \cite{tipo_godel1}. In cylindrical coordinates $(r,\varphi,z)$ the line element is
\begin{equation}
    ds^2= \left[dt + H(r) d\varphi   \right]^2 - D^2(r) d\varphi^2 - dr^2 -dz^2,\label{line}
\end{equation}
where
\begin{align}
    H(r) &= \frac{4\omega}{m^2} \sinh^2 \left(\frac{mr}{2} \right),\label{condition1} \\
     D(r) &= \frac{1}{m} \sinh(mr)\label{condition2}.
\end{align}
with $\omega$ and $m$ being parameters such that $\omega^2 > 0$ and $-\infty \le m^2  \le +\infty$. When $m^2=2\omega^2$ the standard G\"{o}del solution is obtained.

The line element (\ref{line}) can be written as
\begin{equation}\label{type-}
    ds^2 = -dt^2 - 2H(r) dtd\varphi + dr^2 + G(r) d\varphi^2 + dz^2, 
\end{equation}
where $G(r) = D^2(r) - H^2(r)$.  If $G(r) < 0$ in the interval $r_1 < r < r_2$, a circle for $r,t,z = const$ is obtained, i.e., $ds^2 = G(r)d\varphi^2$ denotes a closed time-like curve. For $0 < m^2 < 4\omega^2$ non-causal G\"{o}del circles occur for a region bigger than the critical radius $r_c$. The critical radius is defined by
\begin{equation}
    \sinh^2 \left(\frac{mr_c}{2}\right) \ = \ \left(\frac{4\omega^2}{m^2} -1 \right)^{-1}.
\end{equation}
When $m^2=4\omega^2$ the critical radius is infinity, $r_c=\infty$. This leads to a causal universe. There are no G\"{o}del-type circles, and the breakdown of causality is avoided. 

Using the functions defined in eqs. (\ref{condition1}) and (\ref{condition2}), the metric components are written, explicitly, as
\begin{equation}
     g_{\mu\nu} =\left(\begin{array}{cccc} 
        1 & 0 & \frac{4\,w\,{\mathrm{sinh}^2\left(\frac{m\,r}{2}\right)}}{m^2} & 0\\ 0 & -1 & 0 & 0\\ \frac{4\,w\,{\mathrm{sinh}^2\left(\frac{m\,r}{2}\right)}}{m^2} & 0 & \frac{16\,w^2\,{\mathrm{sinh}^4\left(\frac{m\,r}{2}\right)}}{m^4}-\frac{{\mathrm{sinh}^2\left(m\,r\right)}}{m^2} & 0\\ 0 & 0 & 0 & -1 
    \end{array}\right).
\end{equation}
Then the non-zero Einstein tensor components are
\begin{align}
    G_{00} &= 3\omega^2 - m^2, \\
    G_{02} &=-\frac{4\,w\,{\mathrm{sinh}^2\left(\frac{m\,r}{2}\right)}\,\left(m^2-3\,w^2\right)}{m^2},\\
    G_{11} &= \omega^2, \\
    G_{20} &= -\frac{4\,w\,{\mathrm{sinh}^2\left(\frac{m\,r}{2}\right)}\,\left(m^2-3\,w^2\right)}{m^2},\\
    G_{22} &= \frac{4\,w^2\,{\mathrm{sinh}^2\left(\frac{m\,r}{2}\right)}\,\left(-3\,m^2\,{\mathrm{sinh}^2\left(\frac{m\,r}{2}\right)}+m^2+12\,w^2\,{\mathrm{sinh}^2\left(\frac{m\,r}{2}\right)}\right)}{m^4},\\
    G_{33} &= m^2-\omega^2.
\end{align}
And the Ricci scalar is $R = 2(m^2-\omega^2)$.

For simplicity, let us choose a new basis such that the metric becomes

\begin{equation}
    ds^2 = \eta_{AB} \theta^A \theta^B = (\theta^0)^2 - (\theta^1)^2 - (\theta^2)^2 - (\theta^3)^2, \label{frame}
\end{equation}
where $\eta_{AB}=diag(1, -1,-1, -1)$ and $  \theta^A = e^A\ _{\mu}\ dx^\mu$. Then
\begin{align}
    \theta^{(0)} &= dt + H(r)d\varphi, \\ \theta^{(1)} &= dr, \\ \theta^{(2)} &= D(r)d\varphi, \\ \theta^{(3)} &= dz,    
\end{align}
with $ e^A\,_\mu $ being the tetrad, which the non-null components are
\begin{equation}
        e^{0}\ _{(0)} = e^{1}\ _{(1)} = e^{3}\ _{(3)} = 1, \quad e^{0}\ _{(2)} = - \frac{H(r)}{D(r)}, \quad e^{2}\ _{(2)} = D^{-1}(r).
\end{equation}
Then the non-vanishing components of the Einstein tensor, in the flat (local) space-time, take the form
\begin{align}
     G_{(0)(0)} &= 3\omega^2-m^2, \\
    G_{(1)(1)} &=G_{(2)(2)}= \omega^2,\\
    G_{(3)(3)} &= m^2-\omega^2.
\end{align}
where $G_{AB}=e^\mu_A e^\nu_B G_{\mu\nu}$ has been used.

Then the field equation,  eq.(\ref{mod3}),  in the flat space-time becomes
\begin{equation}\label{mod3_flat}
    \frac{1}{\beta} \left[f'_R R_{AB} - \left(\nabla_A\nabla_B -\eta_{AB} \Box \right) f'_R - \frac{1}{2}\eta_{AB} f \right] = -K\eta_{AB} + 2K_x\partial_A \phi \partial_B \phi+ T_{AB}. 
\end{equation}
Here the cosmological constant is taken as zero. Note that, as the Ricci scalar takes a constant value, the second term on the left hand side of equations (\ref{mod3_flat}) vanishes. The trace of this equation is
\begin{equation}
    f'_R R -2f = \beta\left[-4K + 2K_x \partial^\alpha \phi \partial_\alpha \phi + T \right].
\end{equation}
Defining $\gamma = -4K + 2K_x \partial^\alpha \phi \partial_\alpha \phi$, we get
\begin{equation}\label{trace_flat}
    f'_R R -2f = \beta\left[\gamma + T \right].
\end{equation}
And the  eq.(\ref{mod3_flat}) is written as
\begin{equation}\label{mod32}
    f'_R R_{AB} - \frac{1}{2} \eta_{AB} f = \beta \left[\sigma_{AB} + T_{AB} \right], 
\end{equation}
where $\sigma_{AB} \equiv -K\eta_{AB} + 2K_x\partial_A \phi \partial_B\phi$.

Combining eqs. (\ref{trace_flat}) and (\ref{mod32}) the field equation takes the form
\begin{equation}\label{mod4}
    f'G_{AB} =  \beta \left[\sigma_{AB} + T_{AB} \right] -\frac{1}{2} \left(f + \beta \gamma + \beta T \right)\eta_{AB}.
\end{equation}
Now, let us analyze this field equation for different contents of matter.

\subsection{Perfect fluid}

Here let us consider the perfect fluid as the matter content such that its energy-momentum tensor is
\begin{equation}
    T_{AB}=(\rho+p)u_A u_B+p\eta_{AB},
\end{equation}
where $u_A = (1,0,0,0)$ is the 4-velocity of the fluid, $p$ and $\rho$ is the pressure and energy density, respectively. The trace of the energy-momentum tensor is $T = \rho - 3p$.
Then the components of the field equations for G\"{o}del-type metric are
\begin{align}
    2f'_R (3\omega^2 - m^2) + f &= \beta \left(2\sigma_{00} - \gamma + \rho + 3p   \right)\label{eq_1},\\
    2f'_R \omega^2 - f &= \beta \left(  2\sigma_{11} + \gamma + \rho - p \right)\label{eq_2},\\
    2f'_R(m^2 - \omega^2) - f &= \beta \left( 2\sigma_{33} + \gamma + \rho - p \right)\label{eq_3}.
\end{align}
With $\sigma_{11}=\sigma_{22}=\sigma_{33} = K(\phi,X)$ and $\sigma_{00} = -K(\phi,X) + 2 K_x \dot{\phi^2}$. 

From eqs. (\ref{eq_2}) and (\ref{eq_3}), we get
\begin{equation}\label{sol_godel1}
    f'_R (4\omega^2 - 2m^2) = 0.
\end{equation}
In order to avoid ghost-like and instability, it is considered that $\omega > 0$  and $f'_R > 0$ \cite{Pogo, Nun, Amend}. Then eq. (\ref{sol_godel1}) gives $m^2 = 2\omega^2$, which is the condition for G\"{o}del metric solution. Thus, the field equations are reduced to
\begin{align}
    f'_R m^2 + f &= \beta \left(2\sigma_{00} - \gamma + \rho + 3p \right),\label{sol_type_godel1}\\
    f'_R m^2 - f &= \beta(2\sigma_{33} + \gamma + \rho - p).\label{sol_type_godel2} 
\end{align}
The eqs. (\ref{sol_type_godel1}) and (\ref{sol_type_godel2}) lead to the relations
\begin{align}
    p &= \gamma + \frac{1}{2} \left(\frac{f}{\beta} - \sigma_{00} + \sigma_{33} \right),\label{sol_type_godel3}\\
    \rho &= \frac{f'_R m^2}{\beta} - \gamma -\frac{1}{2} \left(\frac{f}{\beta} - \sigma_{00} + 3\sigma_{33} \right).\label{sol_type_godel4}
\end{align}
From these results, the critical radius in the framework of $f(R,\phi,X)$ theory is
\begin{equation}
    r_c = \frac{2\sinh^{-1}(1)}{\sqrt{\frac{1}{f'_R}\left[\beta\left( \rho + \frac{1}{2} \sigma_{00} + 3\sigma_{33} + \gamma \right) + f\right]}}.
\end{equation}
This shows that it exists non-causal G\"{o}del circles. In addition, the critical radius depends on the $k$-essence theory, $f(R)$ function and matter content. An important note, the expression for the critical radius holds for any $K(\phi,X)$ function and any $f(R)$ theory. 

In order to obtain causal G\"{o}del-type solutions, different sources of matters will be considered. Let us consider as matter content: a combination of a perfect fluid and an electromagnetic field and a single electromagnetic field.

\subsection{ Perfect Fluid Plus Electromagnetic Field}

Here the matter content is composed of a perfect fluid plus an electromagnetic field aligned on $z$-axis and dependent of $z$. For this choice, the non-vanishing components of the electromagnetic tensor in the flat space-time is
\begin{equation}
F_{(0)(3)}=E(z)\quad\mathrm{e}\quad F_{(1)(2)}=B(z),
\end{equation}
where $E(z)$ and $B(z)$ are solutions of Maxwell equations given by
\begin{align}
    E(z)&=E_0\cos\left[2\omega_0(z-z_0)\right],\\
B(z)&=E_0\sin\left[2\omega_0(z-z_0)\right],
\end{align}
with $E_0$ being the amplitude of the electric and magnetic fields and $z_0$ is an arbitrary constant. Thus, the components of energy-momentum tensor of electromagnetic field are 
\begin{equation}\label{energy_sol}
    T_{(0)(0)}^{EM} = T_{(1)(1)}^{EM} = T_{(2)(2)}^{EM} = \frac{E_0^2}{2}, \quad\quad T_{(3)(3)}^{EM} = - \frac{E_0^2}{2}.
\end{equation}
Then the energy-momentum tensor of perfect fluid plus electromagnetic field is 
\begin{equation}
    T_{AB}=(\rho+p)u_A u_B - p\eta_{AB}+T_{AB}^{EM}.
\end{equation}
For this content of matter, the field equations, eq.(\ref{mod4}), are
\begin{align}
    2f'_R (3\omega^2 - m^2) + f -\frac{E_0^2}{2} &= \beta \left(2\sigma_{00} - \gamma + \rho + 3p   \right)\label{eq_1_1},\\
    2f'_R \omega^2 - f + \frac{E_0^2}{2} &= \beta \left(  2\sigma_{11} + \gamma + \rho - p \right)\label{eq_2_2},\\
    2f'_R(m^2 - \omega^2) - f - \frac{E_0^2}{2} &= \beta \left( 2\sigma_{33} + \gamma + \rho - P \right)\label{eq_3_3}.
\end{align}
The eqs. (\ref{eq_2_2}) and (\ref{eq_3_3}) give
\begin{equation}
    m^2 = \frac{E_0^2}{2f'_R} + 2\omega^2.
\end{equation}
By taking the amplitude of the electric field $E_0^2 \ge 0$ and $f'_R > 0$, this equation permits the condition $m^2 = 4\omega$, which implies in $r_c\rightarrow \infty$. Therefore, this solution leads to a causal G\"{o}del-type universe. Moreover, this combination of matter sources does not imply causality violation for any $k$-essence function $K(\phi,X)$ and $f(R)$ theory.

\subsection{Electromagnetic field}

Now a single electromagnetic field aligned on $z$-axis and dependent of $z$ is considered as a matter source. The field equations become
\begin{align}
    2f'_R (3\omega^2 - m^2) + f &= \beta \left(2\sigma_{00} - \gamma    \right)\label{eq_1_1_1},\\
    2f'_R \omega^2 - f &= \beta \left(  2\sigma_{11} + \gamma + 2E_0^2 \right)\label{eq_2_2_2},\\
    2f'_R(m^2 - \omega^2) - f &= \beta \left( 2\sigma_{33} + \gamma  \right)\label{eq_3_3_3}.
\end{align}
The field equations (\ref{eq_2_2_2})  and (\ref{eq_3_3_3}) give rise to the class of G\"{o}del-type solutions
\begin{equation}
    2\omega^2 - m^2 = \beta E_0^2.
\end{equation}
This shows that a causal universe is allowed for the condition $E_0^2 \le 0$. Therefore a causal G\"{o}del-type solution, i.e., $m^2 = 4\omega$ that implies $r_c\rightarrow\infty$, is permitted in $k$-essence theory plus $f(R)$ theory for an electromagnetic field as matter content.

It is interesting note that,  the $f(R,\phi,X)$ theory allows both causal and non-causal G\"{o}del-type solutions.

\section{Conclusions}

Here a hybrid model of $f(R)$ gravity and $k$-essence theory, called $f(R,\phi,X)$ theory, is studied. In this gravitational model the problem of causality is explored. The G\"{o}del and G\"{o}del-type metrics are considered. The G\"{o}del solution is a homogeneous rotating cosmological solution of Einstein equations with pressureless matter and negative cosmological constant, which played an important role in the conceptual development of general relativity. Then is very important to verify the consistency of G\"{o}del solution with alternatives theories of gravity. In addition, G\"{o}del universes could represent a theoretical laboratory for testing modified gravity theories. Here, the consistency between this solution and $f(R,\phi,X)$ gravity theory is developed.  The field equations are solved for different contents of matter. Thus, it is obtained as a result: (i) the G\"{o}del metric is solution in $f(R,\phi,X)$ theory. Then violation of causality is allowed in this theory.  (ii) Considering the G\"{o}del-type metrics with perfect fluid as matter source, the critical radius shows that causality is violated, and depends on the functions $K(\phi,X)$, $f(R)$, $f'_R$ and  of the content of matter. (iii) Looking for a causal solution, the matter source is considered as a combination of a perfect fluid with an electromagnetic field, and simply a single electromagnetic field. For this content of matter, causal regions are permitted. 

\section*{Acknowledgments}

This work by A. F. S. is supported by CNPq projects 308611/2017-9 and 430194/2018-8. J. S. G. thanks CAPES for financial support and Profa. D. O. Maionchi for the fruitful discussions.

\end{document}